\documentclass[traditabstract]{aa}

\usepackage{graphicx}
\usepackage[colorlinks=true,citecolor=blue,linkcolor=blue,urlcolor=blue]{hyperref}
\usepackage{amsmath}
\usepackage{txfonts}
\usepackage{natbib}
\usepackage{upgreek}
\newcommand{\dd}{\mathrm{d}}
\newcommand{\cc}{\mathrm{c}}
\newcommand{\vv}{\mathrm{v}}
\newcommand{\mm}{\mathrm{m}}
\renewcommand{\pi}{\uppi}

\graphicspath{{./figures/}}

\begin{document} 

\title{The influence of mergers on the cluster temperature function and cosmological parameters derived from it}
\titlerunning{Cluster mergers and cosmological parameter constraints}

\author{C.~Angrick \and M.~Bartelmann}

\institute{Zentrum f\"ur Astronomie der Universit\"at Heidelberg, Institut f\"ur Theoretische Astrophysik, Albert-Ueberle-Str.~2, 69120 Heidelberg, Germany\\ \email{angrick@uni-heidelberg.de}}

\date{\emph{A\&A manuscript, version \today}}

\abstract{We develop a parameter-free analytic model to include the effects of mergers into the theoretical modelling of the X-ray temperature function of galaxy clusters. We include this description into our model for the cluster population based on fluctuations of the gravitational potential, which avoids any reference to mass. Comparisons with a numerical simulation reveal that the theoretical model is in good agreement with the simulation results. We show that building the model on the dynamics of spherical rather than ellipsoidal collapse yields better results if emission-weighted temperatures are used, while ellipsoidal collapse yields good agreement between model and simulation for mass-weighted temperatures. Analysing two different samples of X-ray clusters, we quantify the influence of mergers and a conversion between different temperature definitions on the joint determination of $\Omega_\mathrm{m0}$ and $\sigma_8$. If effects of mergers are included, temperature functions based on cluster masses and on the gravitational potential built on spherical collapse are in good agreement with other cosmological probes without any conversion of temperatures.}


\maketitle

\section{Introduction}

Galaxy clusters are a potentially very powerful probe of non-linear cosmological structure formation since their abundance and its evolution depends sensitively on the matter density, the normalisation of density fluctuations and the dark energy. Conventionally, theoretical predictions of the cluster population parametrise clusters by mass. This is potentially problematic since mass is strictly not observable and an integral quantity which, for irregularly shaped bodies without well-defined boundary, is hard to define unambiguously. Calibration relations are needed between the mass and observable quantities such as X-ray temperature and luminosity, which are themselves prone to systematic and random uncertainties.

We have proposed a different approach avoiding any reference to mass \citep{Angrick2009}. The X-ray temperature function of the cluster population, i.e.\ their number-density distribution with X-ray temperature, can be theoretically predicted based on the statistics of gravitational-potential fluctuations. This procedure has several advantages. First, it parametrises the cluster population directly by their temperature, which is a locally defined observable tightly related to the potential depth. Ambiguities caused by the integral definition of the mass are thus avoided. Second, calibration relations for the mass are circumvented, thus removing their scatter from the uncertainty of any inferences \citep[see also][]{Lau2011}. Third, the gravitational potential evolves much less than the matter density, extending the range of validity of linear structure evolution.

We have shown under which conditions this potential-based temperature function reproduces the theoretical predictions based on matter density and mass. Here, we address two subsequent questions. First, we compare the potential-based temperature function to a gas-dynamical, numerical simulation. While we find agreement at low redshift, there is increasing disagreement towards moderate and higher redshifts. This brings us to the development of an analytic model for the effect of cluster mergers on the X-ray temperature function, which leads to very good agreement of our theoretical predictions based on potential statistics with the numerical results. Our analytic model could be considered as providing an analytic complement to the numerical study by \citet{Randall2002}.

Second, we use the potential-based temperature function including the merger model to infer the cosmological parameters $\Omega_\mathrm{m0}$ and $\sigma_8$ from two different samples of galaxy clusters. The results are not conclusive yet, mainly because of tension between observationally inferred temperatures and theoretically motivated temperature definitions, but we find reasonable values for both parameters provided we use a definition of an X-ray temperature function that seems appropriate for the comparison with observational data.

The paper is structured as follows: We first review briefly in Sect.~\ref{sec:tempFunc} the derivation of the potential-based temperature function, extending it to include ellipsoidal rather than spherical collapse. We compare it to a gas-dynamical numerical simulation in Sect.~\ref{sec:simulation} and develop the analytic model for merger effects in Sect.~\ref{sec:mergers}. The inference of cosmological parameters is described in Sect.~\ref{sec:cosmoParam}. Its results are discussed in Sect.~\ref{sec:results}, and we conclude with a summary in Sect.~\ref{sec:summary}.

\section{The potential-based temperature function}
\label{sec:tempFunc}

In the following, we use the X-ray temperature function introduced by \citet{Angrick2009} and an extension thereof based on the generalisation from spherical to ellipsoidal collapse. Both approaches avoid any reference to the globally defined, strictly unobservable \emph{cluster mass}, but are directly derived from the Gaussian statistics of cosmological potential fluctuations.

\subsection{Original form: spherical collapse}
\label{subsec:tempFuncSph}

We briefly sketch the main idea and the basic steps in the derivation of an X-ray temperature function for galaxy clusters that does not invoke cluster mass. It is based on the number density of minima of a homogeneous and isotropic Gaussian random field, discussed in great detail by \citet{Bardeen1986}. For further detail on the derivation, we refer to \citet{Angrick2009}.

The differential number density of potential minima with depth between $\Phi$ and $\Phi+\dd\Phi$ is
\begin{equation}
\label{eq:numDensMinima}
n(\Phi)\,\dd\Phi=\int_{\Delta\Phi_\cc}^\infty\dd(\Delta\Phi)\,\tilde{n}(\Phi,\Delta\Phi)\;\dd\Phi,
\end{equation}
where the number density of relevant potential fluctuations $\tilde{n}(\Phi,\Delta\Phi)$ can be analytically expressed by
\begin{equation}
\label{eq:tildeN}
\tilde{n}(\Phi,\Delta\Phi)=\frac{1}{240\pi^3\sigma_1^3\sqrt{15\gamma}}(F_1+F_2)\exp\left[-\frac{\left(2\sigma_1^2\Delta\Phi+\sigma_2^2\Phi\right)\Phi}{2\gamma}\right]
\end{equation}
with
\begin{align}
\label{eq:F1}
F_1=&\ 2\sigma_2\left(5\Delta\Phi^2-16\sigma_2^2\right)\exp\left[-\frac{\left(6\sigma_0^2\sigma_2^2-5\sigma_1^4\right)\Delta\Phi^2}{2\sigma_2^2\gamma}\right] \nonumber \\ &+\sigma_2\left(155\Delta\Phi^2+32\sigma_2^2\right)\exp\left[-\frac{\left(9\sigma_0^2\sigma_2^2-5\sigma_1^4\right)\Delta\Phi^2}{8\sigma_2^2\gamma}\right], \\
\label{eq:F2}
F_2=&\ 5\sqrt{10\pi}\Delta\Phi\left(\Delta\Phi^2-3\sigma_2^2\right)\exp\left(-\frac{\sigma_0^2\Delta\Phi^2}{2\gamma}\right) \nonumber \\
&\times\left[\mathrm{erf}\left(\frac{\sqrt{5}\Delta\Phi}{2\sqrt{2}\sigma_2}\right)+\mathrm{erf}\left(\frac{\sqrt{5}\Delta\Phi}{\sqrt{2}\sigma_2}\right)\right].
\end{align}
The quantity $\Delta\Phi$ denotes the field's Laplacian and the lower integration boundary is the \emph{critical Laplacian} 
\begin{equation}
\label{eq:critLaplacian}
\Delta\Phi_\cc(a)=\frac{3}{2}H_0^2\Omega_\mathrm{m0}\frac{\delta_\cc(a)}{a},
\end{equation}
where $H_0$ is the Hubble constant, $\Omega_\mathrm{m0}$ is today's matter density with respect to the critical density, $a$ is the scale factor, and $\delta_\cc$ is the critical linear density contrast of the spherical-collapse model. The spectral moments $\sigma_j$ of the redshift-dependent \emph{linear} potential power spectrum $P_\Phi(k,z)$ are defined as
\begin{equation}
\label{eq:specMoments}
\sigma_j^2:=\int_{k_\mathrm{min}}^\infty\dd k\,\frac{k^{2+2j}}{2\pi^2}P_\Phi(k,z)\hat{W}_R^2(k),
\end{equation}
where $\hat{W}_R^2(k)$ is the Fourier transform of a filter accounting for the shape of the gravitational potential of a homogeneous and isotropic overdensity. It is given by
\begin{equation}
\label{eq:filter}
\hat{W}_R(k)=\frac{5\left[3\sin u-u\left(3+u^2\right)\cos u\right]}{2 u^5}
\end{equation}
with $u=kR$ and $R=\sqrt{-2\Phi/\Delta\Phi}$. The cut-off wave-number $k_\mathrm{min}$ introduces an effective sharp high-pass filter in $k$-space and is chosen such that for a given combination $(\Phi,\Delta\Phi)$ the number density $\tilde{n}(\Phi,\Delta\Phi)$ is maximised. This filter step removes long-wave potential-fluctuation modes and thus ensures that clusters are not excluded that move along a potential gradient and thus do not have $\nabla\Phi=0$ at their centres.

We have to relate the cluster's potential $\Phi$ to an observable quantity. This can be relatively easily done for its X-ray temperature in two steps:

\begin{enumerate}

\item The non-linear potential in the cluster centre is given by
\begin{equation}
\label{eq:potCentre}
\Phi_\mathrm{nl}=-2\pi G\rho_\mathrm{b}\delta R^2,
\end{equation}
where $G$ is Newton's constant and $\rho_\mathrm{b}$ the background density of the Universe. It can be related to the linearly evolved potential using the spherical-collapse model for the evolution of the overdensity $\delta$ and the radius $R$ from an initial state at small scale factor to the time of collapse and final virialisation. The relation between the linear and the non-linear potentials, $\Phi_\mathrm{l}$ and $\Phi_\mathrm{nl}$, respectively, is then given by
\begin{equation}
\label{eq:linNonlinPot}
\frac{\Phi_\mathrm{nl}}{\Phi_\mathrm{l}}=\frac{R_\mathrm{ta}}{R_\cc}\frac{\hat{\zeta}^{1/3}}{C}\frac{a_\cc}{a_\mathrm{i}}\frac{D_+(a_\mathrm{i})}{D_+(a_\cc)},
\end{equation}
where
\begin{equation}
\label{eq:C}
C=\frac{3}{5}\left[\hat{\zeta}^{1/3}\left(1+\frac{\Omega_\mathrm{\Lambda,ta}}{\hat{\zeta} \Omega_\mathrm{m,ta}}\right)+\frac{1-\Omega_\mathrm{m,ta}-\Omega_\mathrm{\Lambda,ta}}{\Omega_\mathrm{m,ta}}\right].
\end{equation}
The subscripts `i', `ta', and `c' refer to the initial, turn-around, and collapse times, respectively. $D_+$ denotes the linear growth factor of matter perturbations, and $\hat{\zeta}$ is the overdensity inside the sphere at the time of turn-around. $\Omega_\mm$ and $\Omega_\Lambda$ are the densities of matter and dark energy with respect to the critical density, respectively.

\item The non-linear potential in the centre can be related to the local cluster temperature $T$ using the virial theorem for Newtonian gravity, $\langle E_\mathrm{kin}\rangle=-\frac{1}{2}\langle E_\mathrm{pot}\rangle$, where $\langle E_\mathrm{kin}\rangle=\frac{3}{2}k_\mathrm{B}T$ and $\langle E_\mathrm{pot}\rangle=-\mu m_\mathrm{p}\Phi_\mathrm{nl}$ are kinetic and potential energies, respectively, averaged over a sufficient number of particle orbits. Here, $\mu\approx0.59$ (assuming that the intracluster gas has primordial composition with Helium abundance $Y=0.24$), $k_\mathrm{B}$ is Boltzmann's constant, and $m_\mathrm{p}$ is the proton mass. The central cluster temperature can then be expressed as
\begin{equation}
\label{eq:virTheorem}
k_\mathrm{B}T=-\frac{1}{3}\mu m_\mathrm{p}\Phi_\mathrm{nl}.
\end{equation}
\end{enumerate}
The number density of galaxy clusters as a function of their X-ray temperature is finally given by
\begin{equation}
 \label{eq:numDensT}
n(T)\,\dd T=n(T\stackrel{\mathrm{Eq.~\eqref{eq:virTheorem}}}{\rightarrow}\Phi_\mathrm{nl}\stackrel{\mathrm{Eq.~\eqref{eq:linNonlinPot}}}{\rightarrow}\Phi_\mathrm{l})\left|\frac{\dd\Phi_\mathrm{l}}{\dd\Phi_\mathrm{nl}}\frac{\dd \Phi_\mathrm{nl}}{\dd T}\right|\,\dd T.
\end{equation}

\subsection{Generalisation: ellipsoidal collapse}
\label{subsec:tempFuncEll}

The proposed temperature function is in good agreement with the mass function based on the Press-Schechter formalism \citep{Press1974}, but not with the Sheth-Tormen mass function \citep{Sheth1999,Sheth2002} based on ellipsoidal-collapse dynamics, which better fits mass functions inferred from numerical simulations. Thus, we refined the relation between $\Phi_\mathrm{l}$ and $\Phi_\mathrm{nl}$ by taking deviations from spherical dynamics into account as described in \cite{Angrick2010}. In the following, we present the key ingredients and results of this model, which extends the original work by \citet{Bond1996}.

The dynamics of the dimensionless axes $a_j=R_j/R_\mathrm{pk}$, $j=1, 2, 3$ of a homogeneous ellipsoid, where $R_\mathrm{pk}$ the size of a spherical top-hat corresponding to a mass $M=(4\pi/3)\rho_\mathrm{b}R^3_\mathrm{pk}$, are described by the three coupled differential equations
\begin{equation}
\label{eq:diffEqAxes}
\frac{\dd^2a_j}{\dd a^2}+\left[\frac{1}{a}+\frac{E'(a)}{E(a)}\right]\frac{\dd a_j}{\dd a}+\left[\frac{3\Omega_\mathrm{m0}}{2a^5 E^2(a)}C_j(a)-\frac{\Omega_{\Lambda 0}}{a^2 E^2(a)}\right]a_j=0
\end{equation}
with $C_j=(1+\delta)/3+b_j/2+\lambda_{\mathrm{ext,}j}$. Here, $E(a)$ is the expansion function of a Friedmann model, and the prime denotes the derivative with respect to $a$. The equations are coupled via $\delta=a^3/(a_1 a_2 a_3)-1$, $b_j$, and $\lambda_{\mathrm{ext,} j}$. The latter two are the internal and external shear contributions with
\begin{equation}
\label{eq:intShear}
b_j(t)= a_1(t)a_2(t)a_3(t)\int_0^\infty\frac{\dd\tau}{[a_j^2(t)+\tau]\prod_{k=1}^3[a_k^2(t)+\tau]^{1/2}}-\frac{2}{3}
\end{equation}
and
\begin{equation}
\label{eq:extShear}
\lambda_{\mathrm{ext,}j}(t)=\begin{cases}
  \dfrac{D_+(t)}{D_+(t_\mathrm{i})}\left[
    \lambda_j(t_\mathrm{i})-\dfrac{\delta(t_\mathrm{i})}{3}\right] &\text{(linear approx.)}, \\
  \dfrac{5}{4}b_j(t) &\text{(non-linear approx.)},
  \end{cases}
\end{equation}
respectively, where the subscript `i' here and in the following denotes the initial time. We use a combination of both shear models, called the \emph{hybrid model}, which describes the external shear by a linear approximation until the turn-around of an axis, and then switches smoothly to the non-linear approximation.

The initial conditions for the evolution of the axes are derived from the Zel'dovich approximation and given by
\begin{equation}
\label{eq:initAxes}
a_j(a_\mathrm{i})=a_\mathrm{i}[1-\lambda_j(a_\mathrm{i})],\qquad\left.\frac{\dd a_j}{\dd a}\right|_{a_\mathrm{i}}\approx 1-2\lambda_j(a_\mathrm{i}).
\end{equation}

The initial ellipticity $e_\mathrm{i}$ and prolaticity $p_\mathrm{i}$ of the model are related to the initial overdensity $\delta_\mathrm{i}$ by
\begin{equation}
\label{eq:eandp}
e_\mathrm{i}=\frac{3\sigma}{\sqrt{10\pi}\,\delta_\mathrm{i}}, \qquad p_\mathrm{i}=0,
\end{equation}
where $\sigma^2$ is the variance of the matter power spectrum. These values follow from the probability distribution of the eigenvalues of the Zel'dovich deformation tensor \citep{Doroshkevich1970}.

To stop the collapse, we use the following virialisation conditions for each axis, derived from the tensor virial theorem,
\begin{equation}
 \label{eq:virCondition}
\left(\frac{a_j'}{a_j}\right)^2=\frac{1}{a^2 E^2(a)}\left(\frac{3\Omega_\mathrm{m0}}{2a^3}C_j-\Omega_{\Lambda 0}\right).
\end{equation}

The most important difference compared to the spherical collapse model is the circumstance that the parameters $\delta_\cc$ and $\Delta_\vv$ become mass- or scale-dependent, respectively, and can be larger ($\delta_\cc$ for small masses) or smaller ($\Delta_\vv$ for large masses) by a factor of $\sim2$ compared to the canonical values for the $\Lambda$CDM cosmology. We refer to Fig.~5 of \citet{Angrick2010} for their detailed dependence on mass and redshift.

In the following, we will use the results of the ellipsoidal-collapse model and implement it in our formalism for the X-ray temperature function where results from the spherical collapse were used.

We have to modify Eq.~\eqref{eq:numDensMinima} since in the ellipsoidal-collapse case, the critical Laplacian $\Delta\Phi_\cc$ is now a function of the variable $\Delta\Phi$, which one has to integrate over, through $R=\sqrt{-2\Phi_\mathrm{l}/\Delta\Phi}$ (see Sect.~\ref{subsec:tempFuncSph}). It thus becomes
\begin{equation}
 \label{eq:critLapEll}
n(\Phi_\mathrm{l})\,\dd\Phi_\mathrm{l}=\int_0^\infty\dd(\Delta\Phi)\,\tilde{n}(\Phi_\mathrm{l},\Delta\Phi)\,\uptheta[\Delta\Phi-\Delta\Phi_\cc(\Phi_\mathrm{l},\Delta\Phi)]\,\dd\Phi_\mathrm{l},
\end{equation}
where $\uptheta$ is Heaviside's step function. Note that it is still a good approximation to smooth the density field with an isotropic top-hat of size $R$ and not to introduce an anisotropic smoothing function for the ellipsoidal-collapse model at the initial time, when we are well within the linear regime, and the deviation from sphericity is of order a few times $10^{-5}$. 

The potential in the centre of an ellipsoid is given by
\begin{equation}
 \label{eq:potEllipsoid}
\Phi=-\pi G\rho_\mathrm{b}\delta a_1 a_2 a_3 R_\mathrm{pk}^2\int_0^\infty\frac{\dd\tau}{\sqrt{(a_1^2+\tau)(a_2^2+\tau)(a_3^2+\tau)}}
\end{equation}
\citep[][p.~57]{Binney1987}. For a sphere, $a_1=a_2=a_3=R/R_\mathrm{pk}$, and the integral can be solved analytically yielding $2R_\mathrm{pk}/R$, hence the result for the sphere, Eq.~\eqref{eq:potCentre}, is reproduced. We proceed exactly in the same way as in the previous section, calculating the ratio between linear and non-linear potential at the time of collapse. Again, quantities at a small initial scale factor $a_\mathrm{i}$ are labelled with the index `i' and quantities at the time of collapse with `c'. Using the approximations $a_\mathrm{i,1}\approx a_\mathrm{i,2} \approx a_\mathrm{i,3} \approx a_\mathrm{i}$, and $\Delta_\mathrm{i}\approx 1$, we arrive at
\begin{equation}
 \label{eq:phiLinNonlinEll}
\frac{\Phi_\mathrm{nl}}{\Phi_\mathrm{l}}=\frac{a_\cc}{2\delta_\mathrm{i}}\frac{D_+(a_\mathrm{i})}{D_+(a_\cc)}\int_0^\infty\frac{\dd\tau}{\sqrt{(a_\mathrm{1,c}^2+\tau)(a_\mathrm{2,c}^2+\tau)(a_\mathrm{3,c}^2+\tau)}},
\end{equation}
where we have also used the fact that for the virial overdensity, we can approximate $\delta_\vv=\Delta_\vv-1\approx\Delta_\vv$ since $\Delta_\vv$ is of order 100. All quantities that are necessary to evaluate Eq.~\eqref{eq:phiLinNonlinEll} can be calculated using the ellipsoidal-collapse model by \citet{Angrick2010}. Note that in the ellipsoidal case, the ratio of non-linear and linear potential becomes dependent on both $\Phi_\mathrm{l}$ and $\Delta\Phi$.

To infer an averaged linear potential for a given non-linear one, we marginalise over the dependence on $\Delta\Phi$ weighted by $\tilde{n}(\Phi_\mathrm{l},\Delta\Phi)$ as follows,
\begin{equation}
\langle\Phi_\mathrm{l}\rangle_{\Delta\Phi}(\Phi_\mathrm{nl})=\frac{\int_0^\infty\dd(\Delta\Phi)\,\Phi_\mathrm{l}\,\tilde{n}(\Phi_\mathrm{l},\Delta\Phi)\,\uptheta[\Delta\Phi-\Delta\Phi_\cc(\Phi_\mathrm{l},\Delta\Phi)]}{\int_0^\infty\dd(\Delta\Phi)\,\tilde{n}(\Phi_\mathrm{l},\Delta\Phi)\,\uptheta[\Delta\Phi-\Delta\Phi_\cc(\Phi_\mathrm{l},\Delta\Phi)]},
\end{equation}
where $\Phi_\mathrm{l}=\Phi_\mathrm{l}(\Phi_\mathrm{nl},\Delta\Phi)$ via Eq.~\eqref{eq:phiLinNonlinEll}.

\section{Confronting theory with results from a simulation}
\label{sec:simulation}

We now compare the analytic results for both X-ray temperature functions, using the spherical- and the ellipsoidal-collapse dynamics, to a hydrodynamical simulation by \citet{Borgani2004} for a flat concordance $\Lambda$CDM model with matter density $\Omega_\mathrm{m0}=0.3$, baryon density $\Omega_\mathrm{bar0}=0.04$, Hubble constant $H_0=100~h$~km~s$^{-1}$~Mpc$^{-1}$ with $h=0.7$, and normalisation of the power spectrum $\sigma_8=0.8$ in a box of side-length $192~h^{-1}$~Mpc, starting at redshift $z_\mathrm{start}\simeq46$. The gas physics was implemented using \textsc{gadget-2}, a massively parallel $N$-body/SPH tree code with fully adaptive time-resolution \citep{Springel2005}. The density field was sampled with $480^3$ dark matter and an equal amount of gas particles with masses $M_\mathrm{DM}=6.6\times10^9~\mathrm{M}_\odot$ and $M_\mathrm{gas}=9.9\times10^{8}~\mathrm{M}_\odot$, respectively. During the time evolution, the number of gas particles decreases due to their conversion into star particles, which have slightly lower mass than the gas particles.

The simulation includes radiative cooling processes following \citet{Katz1996}, and a photo-ionising background expected from quasars which ionise the Universe at $z\simeq6$. Star formation is modelled using the hybrid multiphase model for the interstellar medium by \citet{Springel2003}. The simulation code also includes a method to follow the production of metals. However, the effects of metals on the cooling function are not taken into account. This only affects the analysis of simulated galaxy clusters with temperatures $T\lesssim 1$ keV. For the evaluation of the X-ray temperature function, cluster catalogues for the five redshifts $z=0,\ 0.2,\ 0.5,\ 0.7,\ 1$ are available.

\subsection{Problems in the analysis}
\label{subsec:probAnalysis}

In Sect.~\ref{sec:tempFunc}, we established a relation between the depth of a cluster's potential minimum and its temperature in the centre. In the derivation, however, we neglected additional difficulties arising from baryonic physics, such as additional cooling of the gas and feedback processes, e.g.\ from supernovae or AGN. In combination, these effects lead to cool cluster cores. These are in direct contrast to the na{\"\i}ve expectation from Eq.~\eqref{eq:virTheorem} that the temperature in the centre should have a distinct maximum. Moreover, the temperature inferred from the measured cluster spectrum differs from the temperature to be inserted into the virial theorem mainly for the following reasons.
\begin{enumerate}
\item A detector counts the number of photons emitted by the source as a function of their energy. Hence, the more photons are counted in an energy bin, the larger is its weight in the determination of a cluster's temperature. Parts of the cluster having different temperatures are thus weighted by their \emph{emissivity}.
\item The detector response is not the same for all energy bands. Although the numbers of photons collected in two energy bins might be equal, their weight will be different. The detector's \emph{sensitivity} needs to be accounted for.
\end{enumerate}
These two aspects are reflected in the different temperature definitions that are used in numerical simulations. There, the temperature of a cluster is defined as
\begin{equation}
 \label{eq:defineT}
T\equiv\frac{\int W T_\mathrm{gas}\,\dd V}{\int W \,\dd V},
\end{equation}
where $T_\mathrm{gas}$ is the temperature of a gas particle and $W$ is a weight function \citep{Mazzotta2004}. The integral covers a specified volume, e.g.\ the sphere defined by the virial radius. Depending on the choice of the weight function, mainly three different temperatures are used in the literature:
\begin{enumerate}
\item the \emph{mass-weighted} temperature $T_\mathrm{mw}$ with $W=n_\mathrm{gas}$, where $n_\mathrm{gas}$ is the number density of the gas,
\item the \emph{emission-weighted} temperature $T_\mathrm{ew}$ with $W=\Lambda(T_\mathrm{gas})\,n_\mathrm{gas}^2$, where $\Lambda(T_\mathrm{gas})\propto\sqrt{T_\mathrm{gas}}$ is the cooling function,
\item the \emph{spectroscopic-like} temperature $T_\mathrm{sl}$ with $W=n_\mathrm{gas}^2/T_\mathrm{gas}^{3/4}$.
\end{enumerate}
$T_\mathrm{mw}$ is easy to calculate and the temperature to be used in the context of the virial theorem, Eq.~\eqref{eq:virTheorem}. However, it differs from the temperature inferred from observations. Consequently, $T_\mathrm{ew}$ was used to relate temperatures from simulations with spectroscopically derived temperatures. But since disagreements remained, especially due to the inhomogeneous sensitivity of a detector across its energy bands, $T_\mathrm{sl}$ was introduced to match the spectroscopic temperatures of clusters in \textit{Chandra} and \textit{XMM-Newton} surveys better.

For $z=0$, the catalogue includes three differently defined temperatures averaged over various radii. The same is true for the higher redshifts, besides that $T_\mathrm{sl}$ is missing. 

Also based on the simulation by \citet{Borgani2004}, \citet{Rasia2005} provide a fitting formula relating the spectroscopic-like to the emission-weighted temperature for clusters with $T_\mathrm{ew}\gtrsim2$~keV at $z=0$,
\begin{equation}
 \label{eq:TslTew}
T_\mathrm{sl}=(0.70\pm0.01)T_\mathrm{ew}+(0.29\pm0.05).
\end{equation}
We will use that relation in Sect.~\ref{sec:results} later.

In addition to the aforementioned differences in the temperature definitions, hydrodynamical simulations suffer from the \emph{overcooling problem}. The cooling function used in simulations is too efficient in the centres of clusters compared to observations so that more gas can fall into the inner regions than what should be actually allowed. On its way, the gas is adiabatically compressed and heats up. Simulated cluster centres tend to have much higher temperatures than those observed \citep[see][especially their Fig.~9]{Borgani2009}.

A more careful analysis of how to compare properly temperatures that are inferred from both simulations and observations to those used in our theoretical model remains to be done. We postpone it, since first, it requires a detailed review of each individual step in the X-ray analysis of observed clusters, and second, the cluster temperature definitions in the theoretical work might also have to be altered, yielding the possibility to compare them to temperatures inferred from numerical simulations without suffering too much from the overcooling problem. A thorough study of these points will be provided in a forthcoming paper.

\subsection{Comparing different temperature functions}
\label{subsec:propTemp}

Comparing the different temperature definitions averaged inside the virial radius $R_\mathrm{vir}$ with those averaged within $R_{2500}$, where $R_{2500}=(0.12-0.33)\,R_\mathrm{vir}$, i.e.\ only within the inner part of the cluster, where the overcooling problem in the simulations occurs, yields interesting results.  While the emission-weighted temperatures inferred by averaging over the two radii are essentially the same, both the mass-weighted and the spectroscopic-like temperatures differ within $R_\mathrm{vir}$ and $R_{2500}$. At $z=0$, $T_\mathrm{mw}(R_{2500})$ is $\sim$45\% larger than $T_\mathrm{mw}(R_\mathrm{vir})$, and $T_\mathrm{sl}(R_{2500})$ is $\sim$20\% larger than $T_\mathrm{sl}(R_\mathrm{vir})$, showing that $T_\mathrm{ew}$ is most severely biased by overcooling.

\begin{figure}
\resizebox{\hsize}{!}{\includegraphics{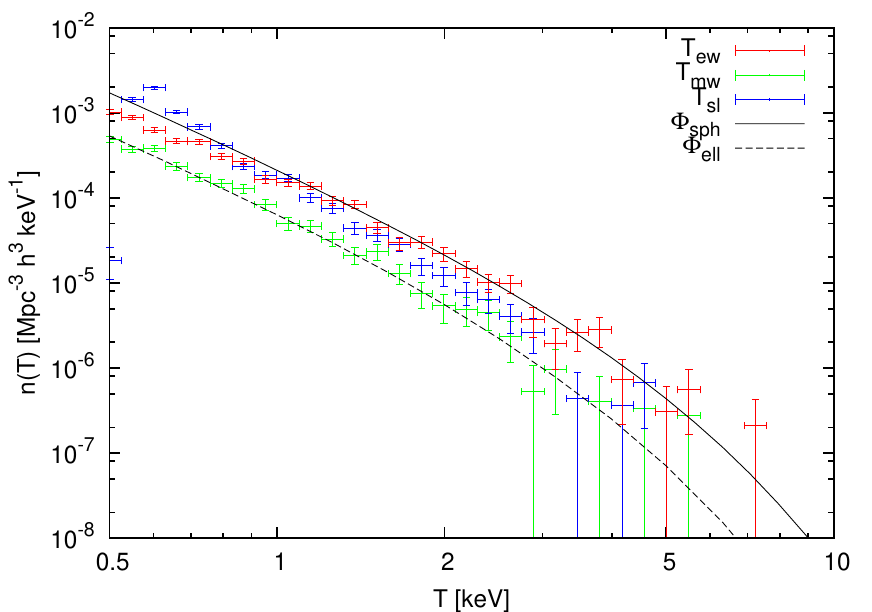}}\\
\resizebox{\hsize}{!}{\includegraphics{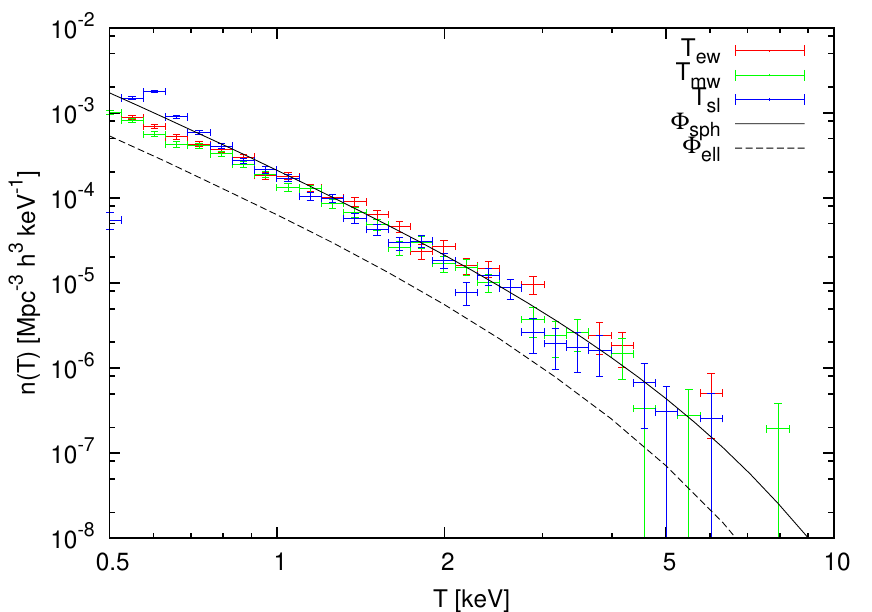}}
\caption{Comparison of the X-ray temperature function inferred from the simulation at $z=0$ using different temperature definitions with the theoretically predicted temperature using either the spherical- ($\Phi_\mathrm{sph}$) or the ellipsoidal-collapse model ($\Phi_\mathrm{ell}$). \emph{Upper panel:} Temperatures evaluated inside $R_\mathrm{vir}$. \emph{Lower panel:} Temperatures evaluated inside $R_\mathrm{2500}$.}
\label{fig:sphEllFunc}
\end{figure}

In Fig.~\ref{fig:sphEllFunc}, we compare the differential number density of clusters inferred from the simulation for $z=0$ using all three different temperature definitions, where the theoretical prediction includes either spherical- or ellipsoidal-collapse dynamics. The temperature function predicted based on ellipsoidal collapse is in better agreement with the simulation if the mass-weighted temperature within the virial radius $R_\mathrm{vir}$ is used (upper panel), while incorporating spherical collapse leads to a theoretical temperature function that agrees well with the simulation if emission-weighted temperatures are used instead.

The reason for this discrepancy is that $T_\mathrm{ew}$ is weighted by $n_\mathrm{gas}^2$ and therefore accentuates the innermost cluster cores which experience overcooling, whereas $T_\mathrm{mw}$ is weighted by $n_\mathrm{gas}$, whence outer parts become more important. The overcooling problem is thus most severe for the temperature function based on $T_\mathrm{ew}$. In this case, the number density of clusters based on the core temperatures is only accidentally well modelled with spherical-collapse dynamics. The weight for the temperature function based on $T_\mathrm{sl}$ causes it to fall between the two predictions.

Interestingly, when considering the temperature functions inferred within $R_{2500}$ (lower panel), i.e.\ restricting the point of view to the very inner part of the halo per definition, the three functions approach each other and are more or less well described by the theoretical description based on the spherical-collapse model, where again the temperature function based on $T_\mathrm{ew}$ fits the prediction best. This behaviour agrees with our earlier remark that the inner parts of clusters that experience overcooling are well described by spherical-collapse dynamics. 

\begin{figure}
\centering
\resizebox{\hsize}{!}{\includegraphics{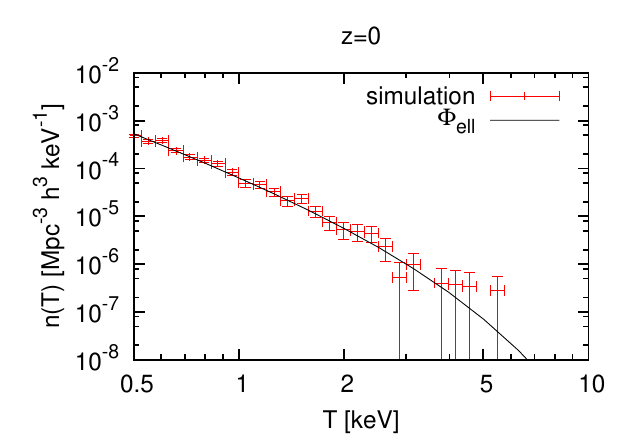} \includegraphics{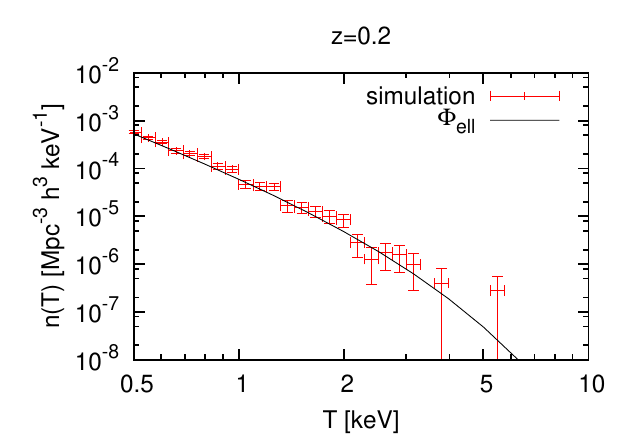}}\\
\resizebox{\hsize}{!}{\includegraphics{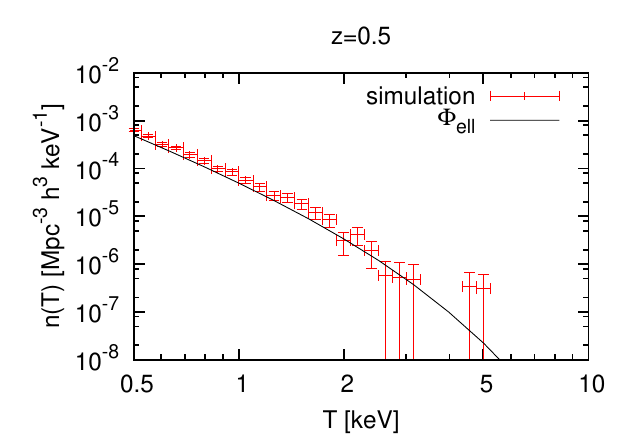} \includegraphics{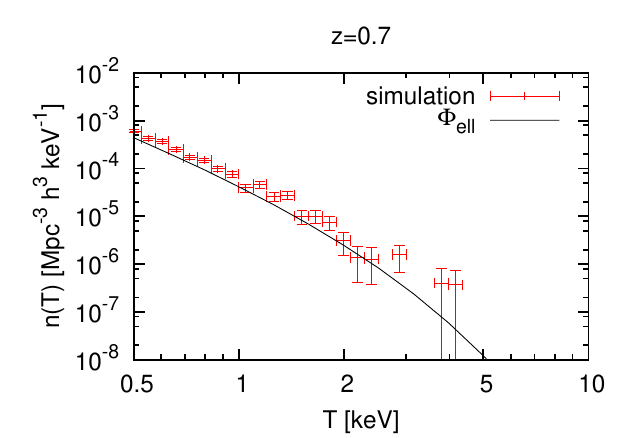}}\\
\resizebox{0.5\hsize}{!}{\includegraphics{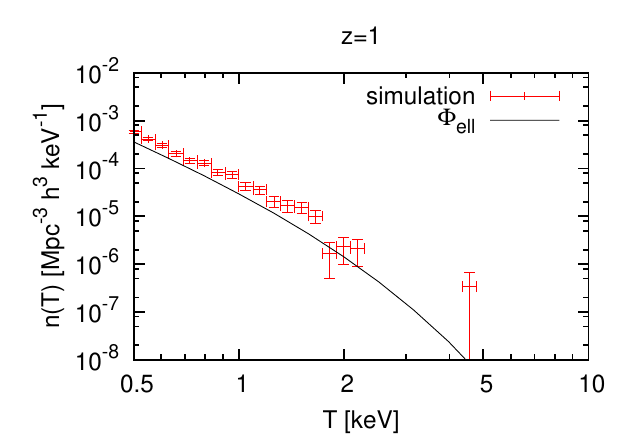}}
\caption{Comparison of the X-ray temperature function from the simulation based on $T_\mathrm{mw}$ within $R_\mathrm{vir}$ with the theoretical prediction including the ellipsoidal-collapse model.}
\label{fig:simPredHighZ}
\end{figure}

Since $T_\mathrm{mw}$ is the temperature that should be used with the virial theorem Eq.~\eqref{eq:virTheorem}, we concentrate on the temperature function based on $T_\mathrm{mw}$ within $R_\mathrm{vir}$ from now on. Figure~\ref{fig:simPredHighZ} compares the results of the simulation based on $T_\mathrm{mw}$ with the potential-based temperature function including ellipsoidal collapse. One can clearly see that the higher the redshift is, the more the results of the simulation disagree from the theoretical prediction. The simulation data systematically fall above the prediction for $z\geq0.5$. In the following section, we hypothesise that this is due to cluster mergers, which are much more numerous in the past, and develop a model for taking them into account in the construction of the cluster temperature function.

\section{Modelling merger effects}
\label{sec:mergers}

\citet{Randall2002} have shown with numerical simulations that mergers do have a strong impact on the X-ray temperature and luminosity functions of haloes and thus, cosmological parameters inferred from them without including this additional effect are biased. Since clusters that are undergoing mergers are shifted from lower to higher temperatures and due to the exponential cut-off at the high-temperature end, especially this part of the temperature function is enhanced. Consequently, \citet{Randall2002} find that the inferred $\sigma_8$ is biased towards higher values, whereas $\Omega_\mathrm{m0}$ is biased towards lower values. Both parameters change by $\sim$15--20\%.

We choose a different approach here, trying to incorporate the essential physical effect of mergers in a simple analytic model based on the extended Press-Schechter formalism by \citet{Lacey1993}. They found that the conditional probability for a halo with mass $M_1$ at time $t_1$ to have mass $M_2+\dd M_2$ at $t_2$ is given by
\begin{equation}
\label{eq:condProb}
\begin{split}
\tilde{p}(S_2,\omega_2|S_1,\omega_1)\,\dd S_2=\ &\frac{1}{\sqrt{2\pi}}\left[\frac{S_1}{S_2(S_1-S_2)}\right]^{3/2}\frac{\omega_2(\omega_1-\omega_2)}{\omega_1} \\
&\times\exp\left[-\frac{(\omega_2S_1-\omega_1S_2)^2}{2S_1S_2(S_1-S_2)}\right]\,\dd S_2,
\end{split}
\end{equation}
where $S_1$ and $S_2$ are the variances of the linear matter power spectrum filtered on scales corresponding to masses $M_1$ and $M_2$, respectively. The $\omega_i$, $i=1, 2$, denote the scaled critical linear density contrasts from the spherical-collapse model at times $t_i$ defined as $\omega_i\equiv\delta_{\mathrm{c},i}(t_i)/D_+(t_i)$. Note that $S_1>S_2$ and $\omega_1>\omega_2$ if defined in this way. Changing variables to $M\equiv M_1$, $\Delta M\equiv M_2-M_1$ and $z\equiv z(\omega_2)$, $\Delta z\equiv z(\omega_1)-z(\omega_2)$ yields
\begin{equation}
 \label{eq:condProbNew}
\begin{split}
p(M,\Delta M,z,\Delta z)\,\dd(\Delta M)\equiv&\ p[S(M+\Delta M),\omega(z)|S(M),\omega(z-\Delta z)] \\
&\times \left|\frac{\dd S}{\dd(\Delta M)}(M+\Delta M)\right|\,\dd(\Delta M).
\end{split}
\end{equation}

Assume now that the temperature increase $\Delta T(M,\Delta M)$ due to a merger of a mass $M$ with another mass $\Delta M<M$ originates from the kinetic energy of the gas transported with the infalling clump, which is completely transformed to thermal energy. The gain of energy is therefore
\begin{equation}
 \label{eq:kinEnergy}
\Delta E=\frac{1}{2}f_\mathrm{b}\Delta M (\Delta v)^2\stackrel{!}{=}\frac{3}{2}N k_\mathrm{B}\Delta T,
\end{equation}
where $f_\mathrm{b}\Delta M$ is the baryon fraction of the lower-mass halo, $\Delta v$ is the relative velocity of the components, and $N=f_\mathrm{b} M/(\mu m_\mathrm{p})$ is the total number of gas particles in the halo of mass $M$. Note that in this ideal case, the factor $f_\mathrm{b}$ cancels exactly. To guess $\Delta v$, assume that the larger component is at rest, while the halo of mass $\Delta M$ approaches from infinity. In this case, the velocity can be easily calculated by equating potential and kinetic energy. The potential at the surface of a homogeneous ellipsoid is given by
\begin{equation}
\label{eq:potEll}
\Phi(\vec{x})=-\frac{3}{4}\frac{MG}{R_\mathrm{pk}}\left[E_1(a_1,a_2,a_3)-\sum_{j=1}^{3}x_j^2\,E_2(a_1,a_2,a_3,a_j)\right],
\end{equation}
with
\begin{align}
E_1(a_1,a_2,a_3)&=\int_0^\infty\frac{\dd\tau}{\prod_{k=1}^3(a_k^2+\tau)^{1/2}}, \\
E_2(a_1,a_2,a_3,a_j)&=\int_0^\infty\frac{\dd\tau}{(a_j^2+\tau)\prod_{k=1}^3(a_k^2+\tau)^{1/2}}
\end{align}
\citep[][p.~57]{Binney1987}, where we have used that the mass of the ellipsoid is given by
\begin{equation}
\label{eq:massEll}
M=\frac{4\pi}{3}\rho a_1 a_2 a_3 R_\mathrm{pk}^3
\end{equation}
(cf.\ Sect.~\ref{subsec:tempFuncEll}), and $\vec{x}$ denotes a position on its surface. A properly averaged potential $\langle\Phi\rangle$, where $\langle\cdot\rangle$ denotes an average over all directions, can be calculated by introducing \emph{ellipsoidal coordinates} $x_1=a_1\sin\theta\,\cos\phi,\ x_2=a_2\sin\theta\,\sin\phi,\ x_3=a_3\cos\theta$ and averaging over the two angles $\theta$ and $\phi$. This finally yields
\begin{equation}
\label{eq:phiAverage}
\langle\Phi\rangle=-\frac{3}{4}\frac{MG}{R_\mathrm{pk}}\left[E_1(a_1,a_2,a_3)-\frac{1}{3}\sum_{j=1}^3a_j^2E_2(a_1,a_2,a_3,a_j)\right].
\end{equation}
Since $\Delta v=\sqrt{-2\langle\Phi\rangle}$, Eq.~\eqref{eq:kinEnergy} yields
\begin{equation}
k_\mathrm{B} \Delta T=-\frac{2}{3}\frac{\mu m_\mathrm{p}\,\Delta M \langle\Phi\rangle}{M},
\end{equation}
where $\langle\Phi\rangle$ is given by Eq.~\eqref{eq:phiAverage}.
We assume that the time scale for the temperature increase is set by the \emph{sound-crossing time}
\begin{equation}
 \label{eq:soundCross}
t_\mathrm{sc}=\frac{R}{c_\mathrm{s}}\qquad\text{with}\qquad c_\mathrm{s}=\sqrt{\frac{5}{3}\frac{k_\mathrm{B}T}{\mu m_\mathrm{p}}}
\end{equation}
\citep{Randall2002}, where $c_\mathrm{s}$ is the sound speed. To infer a proper radius $R$, assume that the halo's mass is given by $M=(4\pi/3)\rho_\mathrm{b}\Delta_\vv R^3$, where $\Delta_\vv=a^3/(a_1a_2a_3)$ in the ellipsoidal-collapse model, so that
\begin{equation}
R(M)=\left(\frac{3M a_1a_2a_3}{4\pi\rho_\mathrm{cr0}\Omega_\mathrm{m0}}\right)^{1/3},
\end{equation}
where $\rho_\mathrm{cr0}$ is the critical density of the Universe today.

\begin{figure}[t]
\centering
\resizebox{0.9\hsize}{!}{\includegraphics{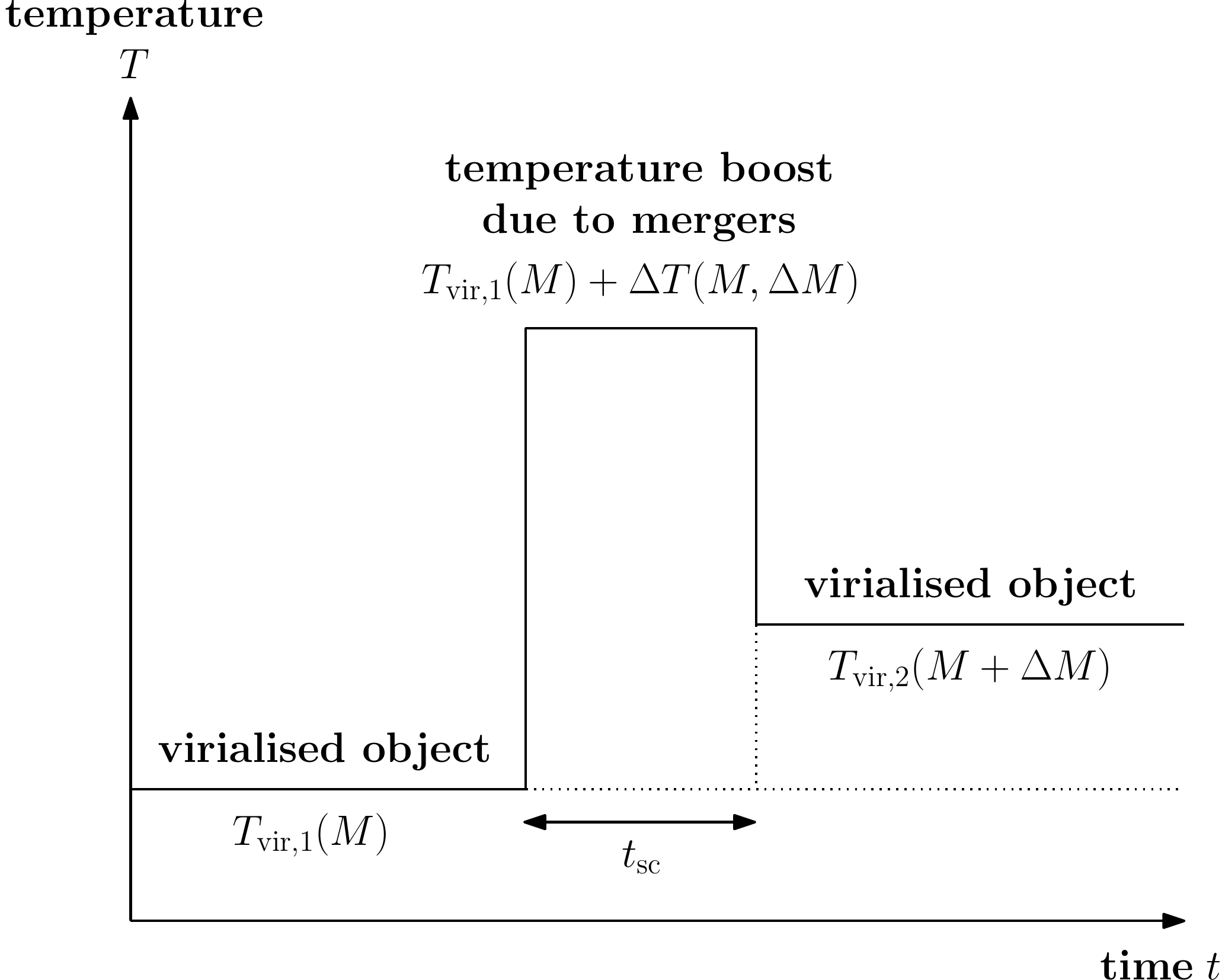}}
\caption{Illustration of a cluster's temperature curve due to a merger as assumed in our model.}
\label{fig:tempBoost}
\end{figure}

As illustrated in Fig.~\ref{fig:tempBoost}, we thus model the temperature boost in an idealised, abrupt way: The halo of mass $M$ has a temperature $T_\mathrm{vir,1}(M)$ before the merger, which increases instantaneously to $T_\mathrm{vir,1}(M)+\Delta T(M,\Delta M)$ for a period $t_\mathrm{sc}$ and then drops instantaneously to $T_\mathrm{vir,2}(M+\Delta M)$, assuming that a virialised halo of mass $M+\Delta M$ has finally formed. To assign a virial temperature $T_\mathrm{vir}$ to a halo of mass $M$, we construct a relation that is based on the combination of Eqs.~\eqref{eq:virTheorem}, \eqref{eq:potEllipsoid}, and \eqref{eq:massEll}, thus assuming a virialised homogeneous ellipsoid. This yields
\begin{equation}
\label{eq:massTempRel}
k_\mathrm{B} T_\mathrm{vir}(M)=\mu m_\mathrm{p}\left(\frac{\Omega_\mathrm{m0}H_0^2 G^2 M^2}{128}\right)^{1/3}\int_0^\infty\frac{\dd\tau}{\prod_{k=1}^3\left[a_k^2(M)+\tau\right]^{1/2}}.
\end{equation}

Starting from the number density of virialised galaxy clusters $n_\mathrm{vir}(T)$, which can be calculated as explained in Sect.~\ref{subsec:tempFuncSph}, we calculate two correction terms. The first is the number density of clusters that reach a temperature $T$ only due to mergers,
\begin{equation}
\label{eq:nPlus}
\begin{split}
 n_+(T)\equiv&\int_0^\infty\dd(\Delta M)\int_0^\infty \dd M\,n[T_\mathrm{vir}(M)]p(M,\Delta M,z,\Delta z) \\
&\times\updelta_\mathrm{D}[T-T_\mathrm{vir}(M)-\Delta T(M,\Delta M)]\,\uptheta(M-\Delta M),
\end{split}
\end{equation}
where we ensure via Dirac's delta function $\updelta_\mathrm{D}$ and Heaviside's step function $\uptheta$ that only combinations of $\Delta M$ and $M$ contribute to the integral for which $T_\mathrm{vir}(M)+\Delta T(M,\Delta M)=T$ and $\Delta M<M$ are
fulfilled. The redshift interval $\Delta z$ is set by the \emph{sound-crossing time}, Eq.~\eqref{eq:soundCross}, since in our simple model, one should be able to see all mergers at redshift $z$ that have happened in the redshift interval $\Delta z$ before. The relation to $t_\mathrm{sc}$ is given by
\begin{equation}
t_\mathrm{sc}=\frac{1}{H_0}\int\limits_z^{z+\Delta z}\frac{\dd z'}{E(z')\,(1+z')}.
\end{equation}

Two assumptions where implicitly made during the derivation: First, the number density $n(T)$ is assumed not to change significantly during $\Delta z$, and second, the increase by $\Delta M$ is only due to a \emph{single} merger event, ignoring multiple simultaneous merger events and smooth accretion. It turns out, however, that $\Delta z$ is short enough for these assumptions not to result in a significant error contribution.

The second correction term arises due to clusters that would have a temperature $T$ if they were virialised, but have a temperature higher than $T$ due to mergers,
\begin{equation}
\label{eq:nMinus}
n_-(T)\equiv\int_0^{M(T)}\dd(\Delta M)\,n(T)p[M(T),\Delta M,z,\Delta z].
\end{equation}
Here, we have to invert Eq.~\eqref{eq:massTempRel} numerically to assign a mass $M$ to the temperature $T$. The total number density of clusters is then given by
\begin{equation}
 \label{eq:nTotal}
n_\mathrm{ges}(T)=n_\mathrm{vir}(T)+n_+(T)-n_-(T).
\end{equation}

\begin{figure}
\centering
\resizebox{\hsize}{!}{\includegraphics{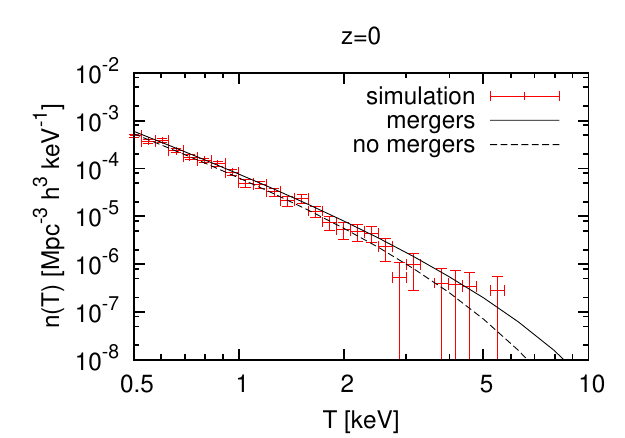} \includegraphics{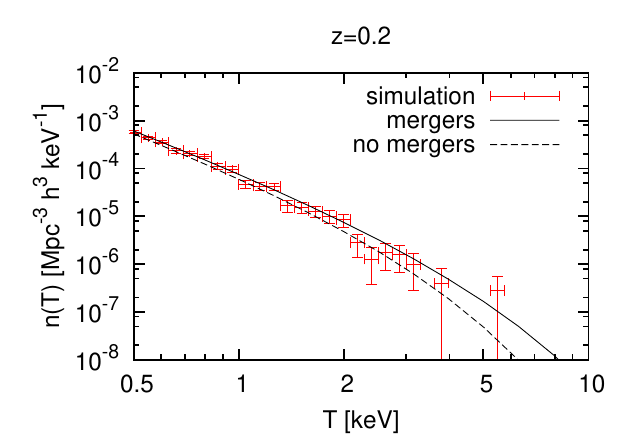}} \\
\resizebox{\hsize}{!}{\includegraphics{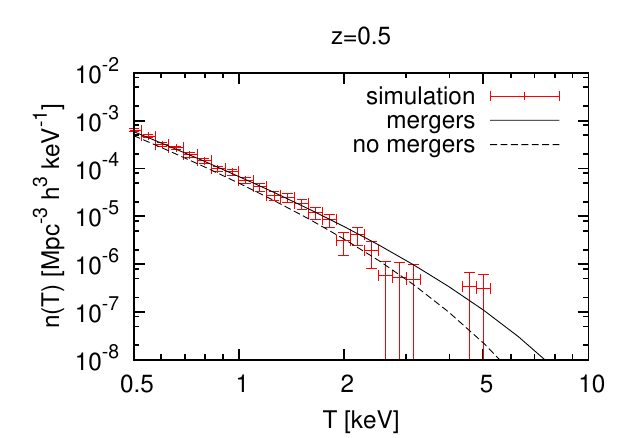} \includegraphics{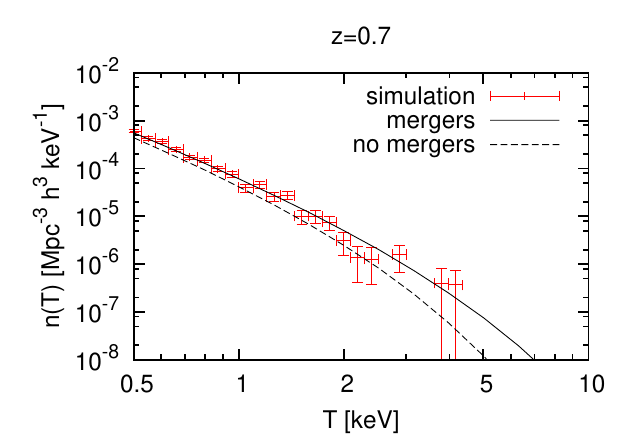}} \\
\resizebox{0.5\hsize}{!}{\includegraphics{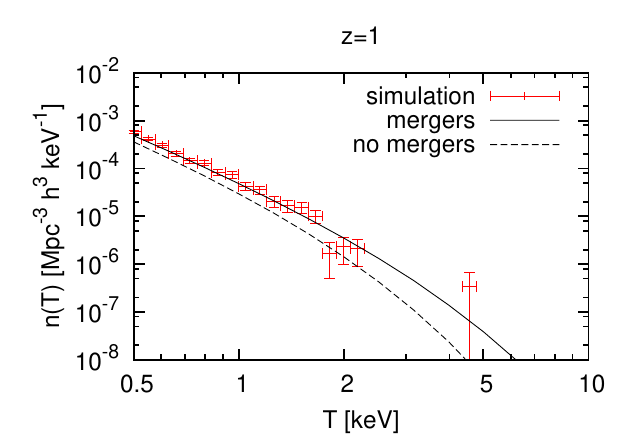}}
\caption{Comparison of the result from the simulation with the theoretical prediction both including (solid line) and excluding (dashed line) the effect of mergers on the temperature function. We compare with the temperature function derived from the simulation that is based on $T_\mathrm{mw}$ within $R_\mathrm{vir}$ (red data points).}
\label{fig:simMerger}
\end{figure}

In Fig.~\ref{fig:simMerger}, we compare the prediction of our X-ray temperature function both with and without the enhancement by mergers with the results from the simulation by \citet{Borgani2004}. The shape of the temperature function influenced by mergers is in qualitative agreement with \citet{Randall2002}. The relative change of the number density increases with temperature and redshift. Using our simple model for the effects of mergers, the temperature function based on the mass-weighted temperature within $R_\mathrm{vir}$ from the simulation is now in good agreement with the theoretical prediction.

\section{Inferring cosmological parameters}
\label{sec:cosmoParam}

In this section, we want to infer the cosmological parameters $\Omega_\mathrm{m0}$ and $\sigma_8$ from two different samples by \citet{Ikebe2002} and \citet{Vikhlinin2009a} using our theoretical model for the X-ray temperature function both with and without the effect of mergers to quantify their influence on the final outcome. Additionally, we compare the results of our approach using the statistics of minima in the cosmic gravitational potential to the traditional method invoking mass functions and an empirical $M$-$T$ relation to see if any of the two gives tighter constraints.

\subsection{The samples}
\label{subsec:data}

The first flux-limited sample by \citet{Ikebe2002} consists of 61 clusters and is based on \textit{ASCA} and \textit{ROSAT} data with a median redshift of $z=0.046$ in the temperature range $1.4~\text{keV}<T<10.55~\text{keV}$. It covers 8.14 steradians, and the flux limit is $1.99\times10^{-11}$~ergs~s$^{-1}$~cm$^{-2}$ in the 0.1--2.4~keV band. The maximal search volume $V_\mathrm{max}$ for each cluster is calculated and listed for an open model with $\Omega_\mathrm{m0}=0.2$ and $\Omega_{\Lambda 0}=0.0$ and for a flat model with $\Omega_\mathrm{m0}=0.2$ and $\Omega_{\Lambda 0}=0.8$. Although $V_\mathrm{max}$ itself depends on the cosmological parameters, it changes only very slightly with them. Neglecting this effect in the following analysis therefore does not induce a significant error.

The second sample encompasses two subsamples by \citet{Vikhlinin2009a}, one at high and one at low redshift, based on \textit{ROSAT} PSPC All-Sky (RASS) and 400~deg$^2$ data. The \emph{low-redshift sample} consists of 49 clusters with flux $f>1.3\times10^{-11}$~erg~s$^{-1}$~cm$^{-2}$ in the 0.5--2~keV band from several samples of RASS with a total area of 8.14 steradians. The redshift coverage is $0.025<z<0.25$ with $\langle z\rangle\approx0.05$, and temperatures are in the range $2.61~\text{keV}<T<14.72~\text{keV}$.

The \emph{high-redshift sample} consists of 36 clusters from the \textit{ROSAT} 400 deg$^2$ survey \citep{Burenin2007} in the redshift range $0.35<z<0.9$ with $\langle z\rangle\approx0.5$ and a redshift-dependent flux limit in the 0.5--2~keV band. For $z>0.473$, the limiting flux is $1.4\times10^{-13}$~erg~s$^{-1}$~cm$^{-2}$, whereas for $z<0.473$, the flux limit corresponds to a minimal X-ray luminosity of $L_\mathrm{X,min}=4.8\times10^{43}(1+z)^{1.8}$~erg~s$^{-1}$. The temperatures of the clusters are in the range $2.13~\text{keV}<T<11.08~\text{keV}$.

For both subsamples, the effective differential search volume $\dd V/\dd z$ as a function of mass $M$ and cosmological parameters $\Omega_\mathrm{m0}$, $\Omega_{\Lambda 0}$ and $h=0.72$ for both subsamples was made available in electronic form on a grid by A.~Vikhlinin. To convert it to a function of temperature, we used the best-fit values of the mass-temperature relation of \citet{Vikhlinin2009a},
\begin{equation}
\label{eq:MT}
M_{500}=M_0\left(\frac{T}{5~\text{keV}}\right)^\alpha E^{-1}(z),
\end{equation}
where $M_0=(3.02\pm0.11)\times10^{14}~h^{-1}~M_\odot$, $\alpha=1.53\pm0.08$, and $E(z)$ is again the expansion function of a Friedmann model.

\subsection{The fitting procedure}
\label{subsec:fit}

Since the errors on the cluster number counts are Poissonian, we use the \emph{$C$ statistic} of \citet{Cash1979} for unbinned data to find the best-fit values for $\Omega_\mathrm{m0}$ and $\sigma_8$, assuming a spatially flat universe, hence $\Omega_{\Lambda 0}=1-\Omega_\mathrm{m0}$. The $C$ statistic is defined as
\begin{equation}
 \label{eq:defC}
C\equiv2\left(N-\sum_i\ln n_i\right),
\end{equation}
where $N$ is the total number of objects expected from the sample assuming a theoretical model and $n_i$ is the theoretically expected differential number density of the $i$-th cluster in the sample with temperature $T_i$ and redshift $z_i$. The sum extends over all sample members.

Although the potential-based temperature function using spherical collapse seems to be in good agreement with the temperature function from the simulation based on $T_\mathrm{ew}$ by \citet{Borgani2004} only because of the overcooling problem, we also include it in our further analysis and either apply a temperature conversion according to Eq.~\eqref{eq:TslTew} or identify the measured temperature directly with the one used in the theoretical framework. In these cases, we simply set $a_1=a_2=a_3=R/R_\mathrm{pk}$ in our merger model.

Thus, we fit in total eight different theoretical models to the two subsamples of \citet{Vikhlinin2009a}:

\begin{enumerate}

\item mass function by \citet{Tinker2008},
\item the same including merger effects,
\item temperature function based on the gravitational potential including spherical-collapse dynamics without temperature conversion,
\item the same including mergers effects,
\item temperature function based on the gravitational potential including spherical-collapse dynamics with temperature conversion,
\item the same including merger effects.
\item temperature function based on the gravitational potential including ellipsoidal-collapse dynamics without temperature conversion,
\item the same including merger effects.

\end{enumerate}

\begin{figure}
 \resizebox{\hsize}{!}{\includegraphics{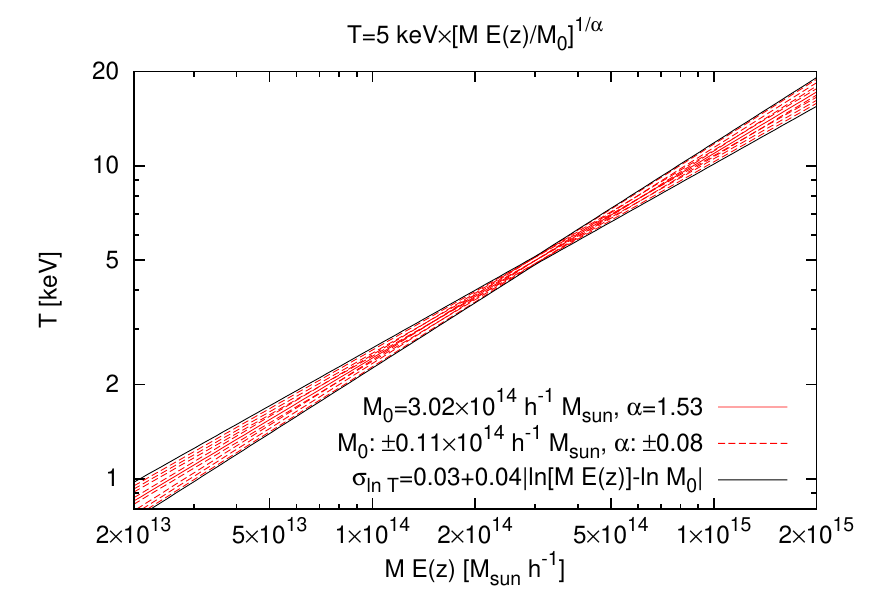}}\\
 \resizebox{\hsize}{!}{\includegraphics{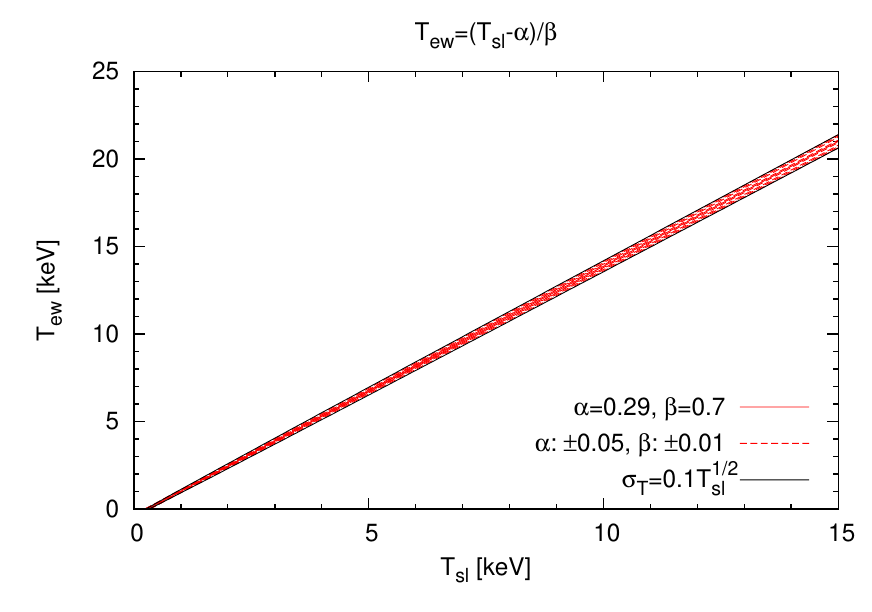}}
\caption{Uncertainties in the mass-temperature relation (\textit{upper panel}) and the $T_\mathrm{sl}$-$T_\mathrm{ew}$ relation (\textit{lower panel}), modelled via varying standard deviations of a log-normal ($\sigma_{\ln T}$) and a normal distribution ($\sigma_T$), respectively.}
\label{fig:standDev}
\end{figure}

In the first two cases, we assume a mass-temperature relation according to Eq.~\eqref{eq:MT} whenever we have to relate a mass to a temperature or vice versa, thus also when applying our analytical merger model. To properly take the scatter into account, we convolve with a log-normal distribution,
\begin{equation}
 \label{eq:logNormalMT}
p_M(T|M)\,\dd T=\frac{1}{\sqrt{2\pi}\,\sigma_{\ln T}T}\exp\left(-\frac{[T-T_0(M)]^2}{2\sigma_{\ln T}^2}\right)\,\dd T
\end{equation}
where $T_0(M)$ is given by Eq.~\eqref{eq:MT} with $M_0=3.02\times10^{14}~h^{-1}~M_\odot$ and $\alpha=1.53$. The standard deviation is mass-dependent and given by
\begin{equation}
\sigma_{\ln T}=0.03+0.04\left|\ln[M E(z)]-\ln M_0\right|
\end{equation}
due to the uncertainty of $M_0$ and $\alpha$; see the upper panel of Fig.~\ref{fig:standDev}. In cases 5 and 6, we convolve with a normal distribution of the form
\begin{equation}
 \label{eq:normalTT}
p_{T_\mathrm{sl}}(T|T_\mathrm{sl})\,\dd T=\frac{1}{\sqrt{2\pi}\,\sigma_T}\exp\left(-\frac{[T-T_\mathrm{ew}(T_\mathrm{sl})]^2}{2\sigma_T^2}\right)\,\dd T,
\end{equation}
where $T_\mathrm{ew}=(T_\mathrm{sl}-0.29)/0.7$ (Eq.~\ref{eq:TslTew}). To take the scatter in the relation into account, the standard deviation is set to $\sigma_T=0.1\sqrt{T_\mathrm{sl}}$; see the lower panel of Fig.~\ref{fig:standDev}. We also use Eq.~\eqref{eq:normalTT} to account for an error contribution in the remaining cases 3, 4 and 7, 8 by simply setting $T_\mathrm{sl}=T_\mathrm{ew}$.

Taking these uncertainties into account, the expected number of objects of each subsample is given by
\begin{equation}
\label{eq:NVikhlinin}
N_\mathrm{low|high}=\int_{z_1}^{z_2}\dd z\int_{T_1}^{T_2}\dd T\,\frac{\dd V_\mathrm{low|high}}{\dd z}(T,z)\int\dd x\, n(x) p_x(T|x),
\end{equation}
where $x$ can be either mass $M$ or spectroscopic-like temperature $T_\mathrm{sl}$ depending on the theoretical model used to fit the data (mass function or temperature function). The integral boundaries depend on the subsample and are given in Sect.~\ref{subsec:data} for $z$ and $T$. The integration over $x$ has to be done over the whole valid range of $p_x$. Finally, the expected differential number density of the $i$-th cluster is simply given by the convolution
\begin{equation}
\label{eq:niVikhlinin}
n_{i,\mathrm{low|high}}=\frac{\dd V_\mathrm{low|high}}{\dd z}(T_i,z_i)\int\dd x\,n(x) p_x(T_i|x).
\end{equation}
To jointly fit both the low and the high-redshift cluster samples of \citet{Vikhlinin2009a}, we have to add the two contributions, finding
\begin{equation}
\label{eq:CVikhlinin}
C_\mathrm{V}=2\left(N_\mathrm{low}-\sum_i \ln n_{i,\mathrm{low}}+N_\mathrm{high}-\sum_j \ln n_{j,\mathrm{high}}\right).
\end{equation}

For the sample by \citet{Ikebe2002}, we proceed analogously, but the situation is much easier since we only deal with one single sample that covers only a small redshift interval. The latter implies that we do not introduce a significant error if we ignore the redshift evolution of the mass or temperature function, respectively, in the analysis. Instead, we compute the theoretical functions at the sample's median redshift of $z=0.046$ in the same way as \citet{Ikebe2002} did so that we only have to integrate over the temperature when calculating the total expected number of objects from the sample. Hence, the $C$ statistic is given by
\begin{align}
\label{eq:CIkebe}
C_\mathrm{I}&=2\left(N-\sum_i \ln n_i\right) \\
\intertext{with}
N&=\int_{T_1}^{T_2}\dd T\, V_\mathrm{max}(T)\int\dd x\,n(x) p_x(T|x), \\
n_i&=V_\mathrm{max}(T_i)\int\dd x\,n(x)p_x(T_i|x).
\end{align}
The conditional probability $p_x(T|x)$ is again given by Eqs.~\eqref{eq:logNormalMT} or \eqref{eq:normalTT}, respectively, thus assuming the same errors on the relations as for the \citet{Vikhlinin2009a} data. To better compare with the results by \citet{Ikebe2002}, we shall also use the classical Press-Schechter mass function instead of the one by \citet{Tinker2008} and relate mass and temperature via Eq.~\eqref{eq:massTempRel} with $a_1=a_2=a_3\equiv R/R_\mathrm{pk}$.

We search for minima of the $C$ statistic as a function of the two cosmological parameters $\Omega_\mathrm{m0}$ and $\sigma_8$, which enter both via $n(x)$ and the volume factors $\dd V/\dd z$ and $V_\mathrm{max}$. Only because the latter is very insensitive to changes in these two parameters, its dependence on $\Omega_\mathrm{m0}$ and $\sigma_8$ can be neglected. \citet{Cash1979} showed that one can create confidence intervals for the $C$ statistic in the same way as it can be done for a $\chi^2$ fit using properties of the $\chi^2$ distribution. Following \citet{Lampton1976}, intervals with confidence $y$ are implicitly given solving
\begin{equation}
 \label{eq:confidence}
y=\int_0^t\dd\chi^2\,f(\chi^2)
\end{equation}
for $t$, where $f$ is the density of the $\chi^2_p$ distribution with $p$ degrees of freedom determined by the number of parameters. For 95\% confidence and $p=2$, it follows that $t=5.991$. Using the minimum of the $C$ statistic, $C_\mathrm{min}$, we can simply calculate the 95\% confidence contours by searching for points in the parameter space for which $C=C_\mathrm{min}+5.991$.

\section{Results}
\label{sec:results}

\begin{figure}
 \resizebox{\hsize}{!}{\includegraphics{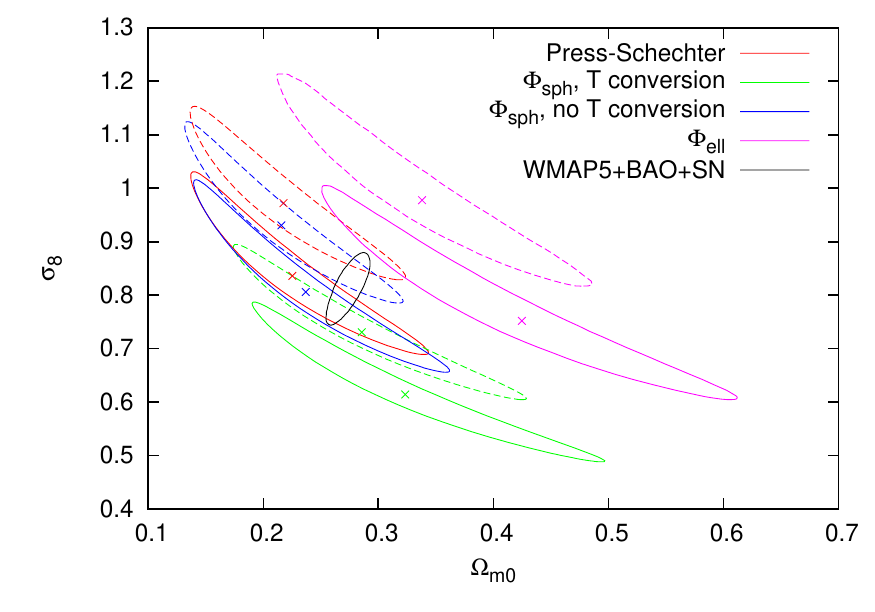}}\\
 \resizebox{\hsize}{!}{\includegraphics{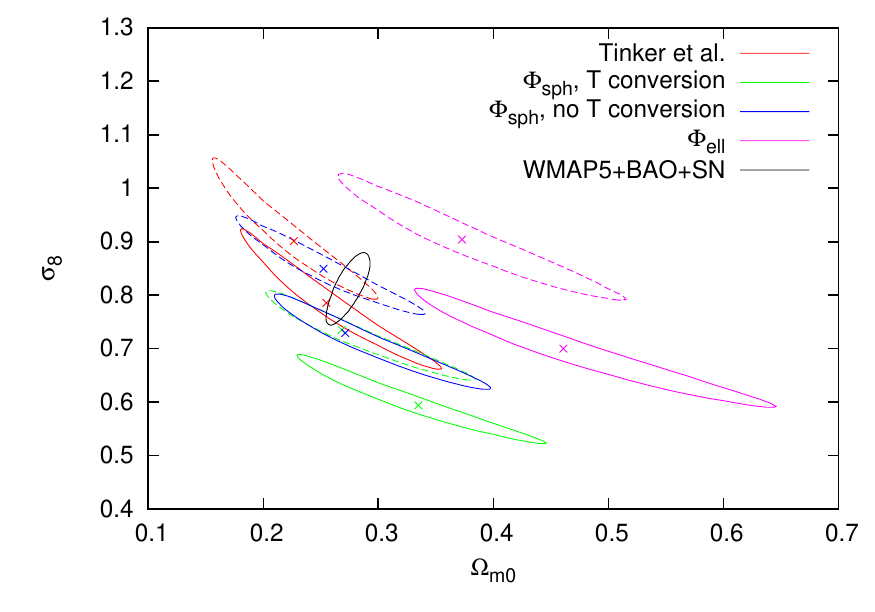}}
\caption{95\% confidence contours for the various theoretical models presented in Sect.~\ref{subsec:fit}. The results for the potential-based temperature function assuming spherical-collapse dynamics are labelled with $\Phi_\mathrm{sph}$, those incorporating ellipsoidal collapse with $\Phi_\mathrm{ell}$, respectively. While the solid curves represent the results of temperature functions that take into account merger effects, the dashed curves are for those which do not account for merging. The crosses mark the respective best-fit values. Parameter constraints from a joint analysis of WMAP5, BAO, and SN data inferred from \citet{Komatsu2009} are indicated by the black solid contour. \textit{Upper panel:} \citet{Ikebe2002} data. \textit{Lower panel:} \citet{Vikhlinin2009a} data.}
\label{fig:confidence}
\end{figure}

In Fig.~\ref{fig:confidence}, we present the 95\% confidence contours for both the \citet{Ikebe2002} and the \citet{Vikhlinin2009a} samples. Comparing the upper and the lower panels, one can see that the results from both data sets are compatible with each other, although pronounced differences exist. However, for the latter, the confidence contours are smaller due to the additional information on the redshift evolution of the temperature function.

The mass-based temperature functions (red contours) and the potential-based temperature functions using spherical collapse without temperature conversion (blue contours) give similar and compatible results, i.e.\ the confidence contours are in agreement with each other and have approximately the same size. When redshift evolution information is added (lower panel), it seems that the direction of degeneracy is slightly changed for the potential-based temperature functions so that $\sigma_8$ can be constrained more tightly compared to the upper panel. For the mass-based temperature function, there is no such effect. Overall, the potential-based functions yield slightly smaller best-fit values for $\sigma_8$ compared to their mass-based counterparts.

If merger effects are taken into account, $\sigma_8$ is significantly lowered while $\Omega_\mathrm{m0}$ is increased for all temperature functions. Consequently, the confidence contours of the results including and excluding mergers do not overlap. This result is in good agreement with the work by \citet{Randall2002}, who also found using numerical simulations that mergers do have a drastic impact on the results for those two parameters.

First converting the measured temperature according to Eq.~\eqref{eq:TslTew} and then comparing it to the theoretical potential-based temperature function incorporating spherical collapse (green contours) has a similar effect as mergers: $\sigma_8$ is decreased while $\Omega_\mathrm{m0}$ is increased. Interestingly, using the potential-based temperature function including merger effects but without temperature conversion (blue solid contours) and using the potential-based temperature function with temperature conversion but \emph{without} merger effects (green dashed contours) give almost identical \citep[as for][]{Vikhlinin2009a} or at least similar \citep[for][]{Ikebe2002} confidence contours. Hence, the effects of temperature conversion and cluster mergers turn out to be highly degenerate. Additionally, merger boosts increase the uncertainty on $\Omega_\mathrm{m0}$ since the confidence contours become more elongated in this direction.

The confidence contours inferred from the potential-based temperature functions built on ellipsoidal-collapse dynamics (magenta contours) are shifted towards higher $\Omega_\mathrm{m0}$ compared to those built on spherical collapse, both including and excluding mergers. This can easily be understood considering Fig.~\ref{fig:sphEllFunc} again. There, the temperature function based on ellipsoidal collapse lies always below the spherical-collapse case for the same cosmological parameters. The reasons are a significantly enlarged $\delta_\cc$ and a smaller $\Delta_\vv$ that leads to a smaller ratio of the non-linearly to the linearly evolved potential. The result of the former is a decrease of the temperature function's amplitude while the latter shifts the curve towards smaller temperatures. Hence, to arrive at a similar fit for a given data set, a higher $\Omega_\mathrm{m0}$ is needed. Additionally, the contours based on ellipsoidal collapse are more extended especially in the $\Omega_\mathrm{m0}$-direction compared to both the spherical case and the mass-based temperature function. This implies that the potential-based temperature function built upon ellipsoidal collapse is the least sensitive to changes in $\Omega_\mathrm{m0}$.

The contours are not in agreement with a joint analysis of the five-year data release of the \textit{Wilkinson Microwave Anisotropy Probe} (WMAP5), and data from baryonic acoustic oscillations (BAO) and supernovae (SN) by \citet{Komatsu2009} (black solid contour). The reason for this discrepancy might be that the ellipsoidal case describes temperature functions very well which are based on the mass-weighted temperature (cf.\ Sect.~\ref{subsec:propTemp}). The latter, however, is different from the temperature that is actually inferred from observations (see Sect.~\ref{subsec:probAnalysis}). A detailed study of this problem is therefore crucial. It will be carried out in a forthcoming paper.

Besides the aforementioned contours from the potential-based temperature function including ellipsoidal collapse, all confidence contours are in agreement with the WMAP5+BAO+SN analysis except one: if both the temperature conversion Eq.~\eqref{eq:TslTew} and merger effects are included in the calculation of the X-ray temperature function (green solid contour), the resulting confidence contour no longer overlaps with that given by \citet{Komatsu2009}.

Although we have assumed in Sect.~\ref{subsec:propTemp} that the simulated temperature function based on $T_\mathrm{ew}$ by \citet{Borgani2004} is only accidentally in good agreement with the potential-based temperature function including spherical collapse because of the overcooling problem in the central parts of simulated clusters, it is remarkable that the spherical function gives results that are compatible with results from both the WMAP5+BAO+SN analysis and classical mass functions. At this point, it is definitely worth analysing whether this is only a coincidence or it really describes temperature functions based on observed temperatures better. The latter would imply that one either has to identify the measured temperature directly with the temperature that is used in the theoretical model and to include additional merger effects, or one has to convert the measured temperature using Eq.~\eqref{eq:TslTew} to a theoretical temperature which is then used in the model. However, in the latter case it seems that one has to exclude merger effects since including both corrections result in values for $\sigma_8$ and $\Omega_\mathrm{m0}$ that are inconsistent with other cosmological probes.

Another reason for the discrepancy between the results from the simulations and the comparison with cluster samples could be that simulated clusters are more elongated than those which are actually observed. The clusters for the sample by \citet{Vikhlinin2009a} for example are selected to appear regular. Furthermore, the different temperature definitions are related in different ways to the potential shape. While $T_\mathrm{mw}$ traces better the cluster potential and hence agrees well with the temperature function built on ellipsoidal collapse (see Fig.~\ref{fig:sphEllFunc}), observed temperatures as well as $T_\mathrm{ew}$ and $T_\mathrm{sl}$ as inferred from simulations follow the more spherical shape of the emitting gas.

The approach including merger effects is more physically motivated since in the theoretical derivation of the pure X-ray temperature function, we explicitly assume virial equilibrium, which only relaxed clusters should have reached. Combining the results of \citet{Randall2002}, who also found that mergers do have a significant impact on the inferred values for $\Omega_\mathrm{m0}$ and $\sigma_8$, with the conclusions of Sect.~\ref{sec:mergers}, we believe that correcting for merger effects should be a necessary step. Note additionally that the scaling relation between both temperatures, Eq.~\eqref{eq:TslTew} was established for clusters at $z=0$ and hence, it is not known how this relation evolves with redshift.

In the preceding discussion, however, we should keep in mind that measurements of $\sigma_8$ from CMB data are degenerate with the optical depth due to reionisation. Breaking this degeneracy requires polarisation data, e.g.~the $T$-$E$ cross-power spectrum. Thus, its value is sensitive to uncertainties in particular in the reionisation parameters and has changed significantly several times with subsequent data releases. Additional information from baryonic acoustic oscillations and type-Ia supernovae do not directly constrain $\sigma_8$ either, but rather tighten constraints on the matter density $\Omega_\mathrm{m0} h^2$ through information on the cosmological constant at fixed spatial curvature. We thus hesitate to accept $\sigma_8$ as derived from WMAP data as a firm reference. Weak-lensing data, that are in principle capable of constraining $\sigma_8$ more directly, still yield a fairly broad range of results, $\sigma_8\sim0.6-0.9$; cf.\ the compilation in \cite{Bartelmann2010}. Some tension between expectations and data are also reflected in the literature. For example, while \citet{Evrard2008} prefer a high normalisation of the power spectrum to be consistent with numerical simulations, \citet{Reiprich2006} concludes that data from the \emph{HIFLUCS} sample prefer a low $\sigma_8$.

\citet{Vikhlinin2009a} take merger effects into account by splitting the clusters of their samples into relaxed and unrelaxed ones by looking at their respective X-ray morphology. If a cluster is classified as unrelaxed, the mass estimate using Eq.~\eqref{eq:MT} is multiplied by a factor of 1.17, assuming that the $M$-$T$ relation for these two cluster samples evolves separately but similarly. This approach is inspired by results of a numerical simulation by \citet{Kravtsov2006}. We think that this rigorous classification of clusters into relaxed and unrelaxed objects is problematic and should be avoided if possible. This can be done using our model of merger effects from Sect.~\ref{sec:mergers}. The resulting solid red contour in the lower panel of Fig.~\ref{fig:confidence} is in good agreement with \citet{Vikhlinin2009b} (see their Fig.~3), where the rigorous classification was made. This and the compatibility with other cosmological probes indicate that our merger model can improve the determination of cosmological parameters from X-ray data without having to decide individually if a cluster is relaxed or not, at least if mass functions and an empirical $M$-$T$ relation are used to model an X-ray temperature function.

\section{Summary \& conclusions}
\label{sec:summary}

In the first part of the paper, we have refined the theoretical X-ray temperature function of \citet{Angrick2009} in two different ways: First, we have used the ellipsoidal-collapse model by \citet{Angrick2010} to account for effects of the dynamics of ellipsoidal rather than spherical collapse and second, we have developed a simple analytic and parameter-free model that takes into account the net effect of temporary X-ray temperature boosts of galaxy clusters that previously underwent mergers on the temperature function. Comparing these two modifications to an $N$-body simulation by \citet{Borgani2004}, we have found the following results:

\begin{itemize}

\item Taking into account ellipsoidal-collapse dynamics is important when comparing the theoretical model to temperature functions of numerical simulations that are based on mass-weighted temperatures averaged over a large volume (e.g.\ inside the virial radius). This is the temperature definition consistent with the virial theorem, Eq.~\eqref{eq:virTheorem}. Temperatures from real observations, however, are similar to the spectroscopic-like temperature $T_\mathrm{sl}$ or the emission-weighted temperature $T_\mathrm{ew}$. We have shown that temperature functions based on $T_\mathrm{ew}$ are in good agreement with the theoretical model only for spherical-collapse dynamics. We have guessed that this is mainly due to the overcooling problem in the centres of simulated clusters and shifts the measured cluster temperatures to larger values. Therefore, it is most prominent for temperatures inferred from the central cluster regions.

\item Especially for $z\gtrsim0.5$, the effects of mergers cannot be excluded since the higher the redshift, the more clusters are unrelaxed and therefore deviate from virial equilibrium. Our simple analytic and parameter-free model based on the merger probability derived by \citet{Lacey1993} can account for these effects. Including it in our theoretical modelling leads to a substantially improved agreement with the simulation for redshifts $0\leq z\leq 1$.

\end{itemize}

In the second part, we have used both mass-based and potential-based X-ray temperature functions including either spherical- or ellipsoidal-collapse dynamics together with samples of \citet{Ikebe2002} and \citet{Vikhlinin2009a} to constrain the cosmological parameters $\Omega_\mathrm{m0}$ and $\sigma_8$. We have analysed the influence of merger effects on the inferred values of both parameters by using our analytical model. In addition, we have tested whether it is necessary to convert the measured temperatures by means of Eq.~\eqref{eq:TslTew} or if it is possible to compare them to the theoretical prediction directly. The main results are the following:

\begin{itemize}

\item The temperature function based on spherical-collapse dynamics by \citet{Angrick2009} leads to confidence contours in the $\Omega_\mathrm{m0}$-$\sigma_8$ plane that are compatible to those inferred from classical mass-based temperature functions. The best-fit values might be shifted by a few percent. However, the formalism does not refer to cluster masses whose relation to observables have to be calibrated.

\item Using the potential-based temperature function that incorporates ellipsoidal collapse as the theoretical model yields confidence contours that are not compatible with a joint analysis by \citet{Komatsu2009} using WMAP5+BAO+SN data since the inferred values for $\Omega_\mathrm{m0}$ seem too large. We believe that the reason is the discrepancy between mass-weighted temperature to be used in the virial theorem and the actual measured temperature that is similar to $T_\mathrm{sl}$, but also selection effects such as preferring more regular clusters over ellipsoidal clusters in a cluster sample as well as different relations of the various temperatures to the potential structure ($T_\mathrm{mw}$ traces more the cluster potential, whereas $T_\mathrm{ew}$ and $T_\mathrm{sl}$ follow mainly the spherical shape of the emitting gas) introduce a bias that can lead to the discrepancy between our analysis based on the temperature function incorporating ellipsoidal collapse and the WMAP5+BAO+SN data.

\item In case of the temperature function built on spherical collapse, identifying the measured temperature directly with the temperature that is used in the theoretical model and additionally including merger effects, or converting the measured temperature using Eq.~\eqref{eq:TslTew} before comparing to the model and disregarding merger effects give similar results for $\Omega_\mathrm{m0}$ and $\sigma_8$. Since the comparison to the simulation and the results from the mass-based temperature functions indicate that merger effects lead to biases, we believe that the direct comparison between measured and theoretical temperatures is the correct choice when using our potential-based X-ray temperature function.

\item Almost all of our results using the potential-based temperature function relying on spherical collapse are compatible with the WMAP5+BAO+SN data. Only if both the temperature conversion Eq.~\eqref{eq:TslTew} and merger effects are taken into account, the resulting confidence contours disagree with the latter. Since merger effects do have a significant impact on the determination of $\Omega_\mathrm{m0}$ and $\sigma_8$ \citep[one of the main results of][]{Randall2002}, a temperature conversion does not seem to be necessary in the context of the potential-based temperature function. The reason why the spherical temperature function gives such good results at all has to be further analysed since the agreement with the data from the simulation of \citet{Borgani2004} is suspected to be mainly due to the overcooling problem in the centre of simulated clusters. A deeper understanding of how to relate measured temperatures to those used in our theoretical models is necessary.

\item The combination of the mass function by \citet{Tinker2008} with our simple analytic merger model yields similar results as the technique by \citet{Vikhlinin2009a,Vikhlinin2009b}, i.e.\ rigorously classifying galaxy clusters into relaxed and unrelaxed objects. However, we believe that our model is physically better justified since we take merger effects into account statistically.

\item Although the results for the samples by \citet{Ikebe2002} and \citet{Vikhlinin2009a} are in agreement with each other, pronounced differences exist, implying that the final confidence regions from our analysis of the data might still be biased from systematics in the X-ray analysis and therefore, even better agreement with other cosmological probes could be achieved.

\end{itemize}

\acknowledgements{We want to thank the anonymous referee for the careful review which helped improving the presentation of our results. CA wants to thank S.~Borgani and F.~Pace for making available simulation data of galaxy cluster temperatures, A.~Vikhlinin for kindly providing tabulated data of survey volumes, and the Deutsche Forschungsgemeinschaft for financial support under grant number BA 1369/12-1. CA is a member of the Heidelberg Graduate School of Fundamental Physics, and the IMPRS for Astronomy \& Cosmic Physics at the University of Heidelberg. MB was supported in part by the Transregio-Sonderforschungsbereich TR\ 33 of the Deutsche Forschungsgemeinschaft.}

\bibliography{mergers.bib}

\end{document}